\documentclass[12pt,a4paper]{article}
\usepackage{amsfonts,latexsym}
\usepackage{graphicx,color}
\usepackage{dcolumn}
\usepackage{graphicx}
\usepackage{amsmath}
\usepackage{amsfonts}
\usepackage{amssymb}

\renewcommand{\thefootnote}{\fnsymbol{footnote}}

\begin{document}

\vspace{12mm}

\begin{center}
{{{\Large {\bf Revisiting  superradiant stability of Kerr-Newman black holes under a charged massive scalar   }}}}\\[10mm]

Yun Soo Myung$^{a,b}$\footnote{e-mail address: ysmyung@inje.ac.kr}\\[8mm]

{${}^a$Institute of Basic Sciences and Department  of Computer Simulation, Inje University Gimhae 50834, Korea\\[0pt]}

{${}^b$Asia Pacific Center for Theoretical Physics, Pohang 37673, Korea}

\end{center}
\vspace{2mm}

\begin{abstract}
We revisit  the superradiant stability of Kerr-Newman black holes under a charged massive scalar perturbation.
We  obtain a newly suitable potential  which is  not singular at the outer horizon when a radial  equation  is expressed  the Schr\"{o}dinger-type equation in terms of the tortoise coordinate.
From the potential analysis, we find   a condition for the  superradiant stability of Kerr-Newman black holes.

\end{abstract}
\vspace{5mm}

\vspace{1.5cm}

\hspace{11.5cm}
\newpage
\renewcommand{\thefootnote}{\arabic{footnote}}
\setcounter{footnote}{0}


\section{Introduction}

For an asymptotically flat Kerr black hole, if the incoming scalar wave has a non-zero mass $\mu$, its mass would act as a natural mirror.
In this case,  one might find a superradiant instability of the black hole when the parameters of black holes and scalar field are in certain parameter spaces~\cite{Zouros:1979iw}.
Here, a trapping well of the scalar potential plays the important role in making superradiant instability
 because superradiant modes are localized in the trapping well. If there is no trapping well, this black hole seems to be  superradiantly stable.

Recently, the superradiant stability of Kerr-Newman (KN) black holes under a charged massive scalar perturbation can be
achieved if two conditions of  $qQ/\mu M> 1$ and $r_-/r_+\le 1/3$ are obtained from the potential analysis~\cite{Xu:2020fgq}, in addition that the superradiance condition and the bound-state condition hold.   Actually, the disappearance of a trapping well is a necessary condition for superradiant stability.

However, the potential determining the above two conditions
seems  to be inappropriate  for analyzing  the superradiant stability because  $V(r\to r_+)$ is singular and thus, it does not lead to  $\omega^2-(\omega-\omega_c)^2$ for imposing the superradiance condition. Furthermore, a defining equation for the potential (written by $r$)  does not look like the Schr\"{o}dinger-type  equation since it was  not written by making use of a tortoise coordinate $r_*$. The importance of $r_*$ arises from the fact that its range from $-\infty$ to $\infty$ exhausts the entire part of spacetime which is accessible to an observer outside the outer horizon~\cite{Chan:1983}. If one uses $r$-coordinate, it covers a small region of $r\in[r_+,\infty)$ only.

  Importantly,  we wish to point out that  such a potential was originated from the derivation of a scalar equation expressed by $r$ [Eq.(15) in Ref.~\cite{Hod:2012zza}] where a famous condition for  a trapping well was derived as $\mu^2/2<\omega^2<\mu^2$ (or, $\mu/\sqrt{2}<\omega<\mu)$ in the study of  Kerr black hole under a massive scalar perturbation.  A potential derived in this way was used to determine the superradiant instability regime of the KN black hole~
  \cite{Huang:2016qnk}.
   Also, a similar  potential was  employed  subsequently  in deriving the conditions for superradiant stability of Kerr black holes~\cite{Huang:2019xbu}. Very recently, a similar approach was applied to testing  the extremal rotating  black holes  under a charged massive scalar perturbation~\cite{Lin:2021ssw}, dyonic Reissner-Nordstr\"{o}m (RN) black holes under a charged massive scalar perturbation~\cite{Zou:2021mwa},  higher-dimensional non-extremal RN
black holes under a massive scalar perturbation~\cite{Huang:2021jaz}, and  D-dimensional extremal
RN black holes under a charged
massive scalar perturbation~\cite{Huang:2022nzm}. It may  not be  valid  to adopt such potentials to analyzing the superradiant (in)stability of a massive scalar propagating  around rotating black holes.

In this work, we wish to revisit  the superradiant stability of KN black holes under a charged massive scalar perturbation.
We  obtain an appropriate  potential $V_{KN}(r\to r_+)\sim \omega^2-(\omega-\omega_c)^2$  when writing the Schr\"{o}dinger-type equation in terms of the tortoise coordinate $r_*=\int [(r^2+a^2)/\Delta]dr$. For $Q=0$, this reduces to the well-known potential for a massive scalar perturbation propagating  around the Kerr black hole background~\cite{Zouros:1979iw}.
From the analysis based on $V_{KN}(r)$, we could not  derive the two conditions of $qQ/\mu M> 1$ and $r_-/r_+\le 1/3$  for the  superradiant stability of a  charged massive scalar propagating around the  KN black holes. However, we obtain a condition for the superradiant stability.

\section{A charged massive scalar on the KN black holes }
First of all, we introduce the
Boyer-Lindquist coordinates to represent the  KN
black hole  with mass $M$, charge $Q$,  and angular momentum $J$
\begin{eqnarray}
ds^2_{\rm KN}&=&\bar{g}_{\mu\nu}dx^\mu dx^\nu \nonumber \\
&=&-\frac{\Delta}{\rho^2}\Big(dt -a \sin^2\theta d\phi\Big)^2 +\frac{\rho^2}{\Delta} dr^2+
\rho^2d\theta^2 +\frac{\sin^2\theta}{\rho^2}\Big[(r^2+a^2)d\phi -adt\Big]^2 \label{KN}
\end{eqnarray}
with
\begin{eqnarray}
\Delta=r^2-2Mr+a^2+Q^2,~ \rho^2=r^2+a^2 \cos^2\theta,~{\rm and}~a=\frac{J}{M}.
 \label{mps}
\end{eqnarray}
In addition, the electromagnetic potential is given by
\begin{equation}
\bar{A}_\mu=\frac{Q r}{\rho^2}\Big(-1,0,0, a\sin^2\theta\Big).
\end{equation}
The outer and inner horizons are determined by imposing $\Delta=(r-r_+)(r-r_-)=0$ as
\begin{equation}
r_{\pm}=M\pm \sqrt{M^2-a^2-Q^2}.
\end{equation}
So, it is clear that $\Delta \to 0$, as $r\to r_\pm$.

A charged massive scalar perturbation $\Phi$  on the background of KN black holes is described  by
\begin{equation}
(\bar{\nabla}^\mu-i q \bar{A}^\mu)(\bar{\nabla}_\mu-i q \bar{A}_\mu)\Phi-\mu^2\Phi=0.\label{phi-eq1}
\end{equation}
Reminding the axis-symmetric
background (\ref{KN}), it is convenient to separate the scalar perturbation
into modes
\begin{equation}
\Phi(t,r,\theta,\phi)=\Sigma_{lm}e^{-i\omega t + i m \phi} S_{l m
}(\theta) R_{l m}(r)\,, \label{sep}
\end{equation}
where $S_{\ell m}(\theta)$ is spheroidal harmonics with $-m\le \ell
\le m$ and $R_{l m}(r)$ satisfies a radial part of the wave
equation. Plugging (\ref{sep}) into (\ref{phi-eq1}), one has  the
angular equation  for $S_{l m}(\theta)$ and the radial Teukolsky equation for $R_{l m}(r)$ as~\cite{Hod:2014baa}
\begin{eqnarray}
&& \frac{1}{\sin \theta}\partial_{\theta}\Big(
\sin \theta
\partial_{\theta} S_{\ell m}(\theta) \Big )+ \left [\lambda_{lm}+ a^2 (\mu^2-\omega^2) \sin^2
{\theta}-\frac{m^2}{\sin ^2{\theta}} \right ]S_{l m}(\theta) =0,
\label{wave-ang1}
\end{eqnarray}
\begin{eqnarray}
\Delta \partial_r \Big( \Delta \partial_r R_{\ell m}(r) \Big)+U(r)R_{lm}(r)=0,
\label{wave-rad}
\end{eqnarray}
where
\begin{eqnarray}
U(r)=[\omega(r^2+a^2)-am-qQr]^2+\Delta[2am\omega-\mu^2(r^2+a^2) -\lambda_{lm}]. \label{u-pot}
\end{eqnarray}
Eq.(\ref{wave-rad}) could be used directly for computing absorption cross-section and quasinormal modes of the scalar, and scalar clouds  on the background of KN black holes.
\begin{figure*}[t!]
   \centering
  \includegraphics{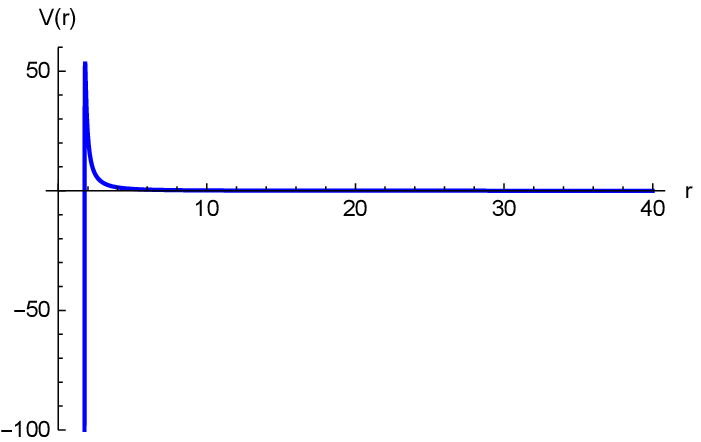}
  \hfill%
  \includegraphics{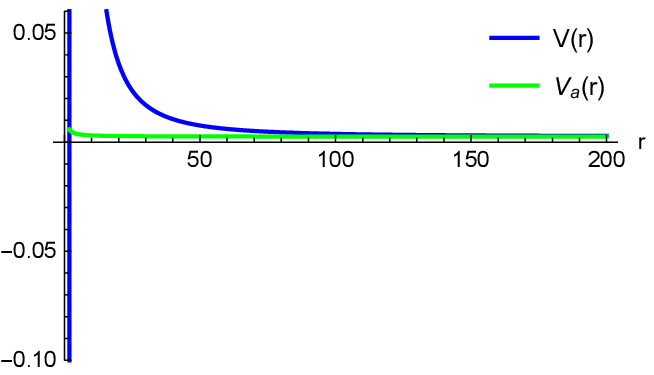}
\caption{(Left) Potential $V(r)$ as function of $r$ with $M=1,Q=0.6,\omega=0.02,a=0.3,m=1,q=0.1,\lambda=12,\mu=0.05$  for $r\in[r_+=1.741,40]$.
The potential is negatively singular ($-1.324\times 10^{13}$) at $r=r_+$. (Right) Potential $V(r)$ and its asymptotic potential $V_a(r)$ as functions of $r\in[r_+,200]$. $V_a(r)$ mimics $V(r)$   for very large  $r\ge 150$.}
\end{figure*}

Introducing $\Psi_{lm}(r)=\sqrt{\Delta}R_{lm}(r)$, one finds easily that  the radial equation  (\ref{wave-rad}) leads to
\begin{equation}
\frac{d^2\Psi_{lm}(r)}{dr^2} +\Big[\omega^2-V(r)\Big]\Psi_{lm}(r)=0,\label{wave-new}
\end{equation}
where an effective potential $V(r)$ is given by~\cite{Huang:2016qnk,Xu:2020fgq}
\begin{eqnarray}
V(r)=\omega^2&-&\frac{1}{\Delta^2}\Big[\Big(\omega(r^2+a^2)-am-qQr\Big)^2 \nonumber  \\
 &+& \quad \quad\Delta\Big(2am\omega-\mu^2(r^2+a^2) -\lambda_{lm}\Big)+M^2-a^2-Q^2\Big]. \label{f-pot}
\end{eqnarray}
We note that $V(r)$ was obtained just by imposing on the disappearance of $\Psi'_{lm}$ in Eq.(\ref{wave-new}).
Here, it is important to point out that  Eq.(\ref{wave-new}) is not a proper Schr\"{o}dinger-type (one-dimensional) equation expressed in terms of a tortoise  coordinate $r_*$.
As is shown in (Left) Fig. 1, $V(r)$ is not a suitable potential to analyze the superradiant stability  because $V(r\to r_+)$ is negatively singular like `$-1.324\times 10^{13}$'.
This arises  because  $\Delta\to 0$ as $r\to r_+$ in the denominators.
Instead, to find the superradiance condition,  one should  have $V(r\to r_+)\sim \omega^2-(\omega-\omega_c)^2$ where $\omega_c$ the critical frequency is given by
\begin{equation}
 \omega_c=m\Omega_H+q\Phi_H \label{omegac}
 \end{equation}
with $\Omega_H=a/(r_+^2+a^2)$ and $\Phi_H=Qr_+/(r_+^2+a^2)$.

To obtain a condition for superradiant instability (trapping well), it is necessary to introduce an asymptotic form  $V_a(r)$.
It was proposed that a trapping well exists if its first derivative must be positive ($V'_a(r)>0$)~\cite{Hod:2012zza}.
On the other hand, a trapping well does not exist if its first derivative must be negative ($V'_a(r)<0$)~\cite{Xu:2020fgq}.
In this case, one finds $V_a(r)$ when expanding $V(r)$ for very large $r$ as
\begin{equation}
V(r\to \infty)\to V_a(r)+{\cal O}\Big(\frac{1}{r^2}\Big), \label{vasym}
\end{equation}
where
\begin{equation}
V_{a}(r)=\mu^2+\frac{2M\mu^2+2q Q \omega-4M \omega^2}{r}. \label{vasym-p}
\end{equation}
Here, its first derivative takes the form
\begin{equation}
V_{a}'(r)=-\frac{2M\mu^2+2q Q \omega-4M\omega^2}{r^2}. \label{vasym-d}
\end{equation}
It was found that a condition for no trapping well leads to  $(V_a'(r)<0 \to 2M\mu^2+2q Q \omega-4M\omega^2 >0)$, which implies one condition of  $ qQ/\mu M>1$.
However, from  (Right) Fig. 1,   $V_a(r)$  may  represent $V(r)$ well for very large $r\ge 150$. So, the bound of  $ qQ/\mu M>1$ might  not be  regarded as   a necessary condition for no trapping well.
It suggests that the asymptotic potential  $V_a(r)$ should include  ${\cal O}(1/r^2)$-terms  to represent $V(r)$ for large  $r\ge20$ appropriately. The other condition of $r_-/r_+ \le 1/3$ was derived  from the potential analysis under the condition of $ qQ/(\mu M)>1$. However, this expression should be replaced by
\begin{equation}
\frac{r_-}{r_+} \le \frac{1}{2\sqrt{\Big(\frac{a}{r_+}\Big)^2+1} -1}.
\end{equation}
 Up to now, we have briefly explained how two conditions for superradiant stability are derived from the analysis of potential $V(r)$  in (\ref{f-pot}).
 At this stage, we wish to mention that all analyses based on $V(r)$ may lead to the undesirable  results.
 
\section{Superradiance analysis with  a new potential}
Let us introduce  the tortoise  coordinate $r_*$  implemented by $dr_*=
\frac{r^2+a^2}{\Delta}dr$ to derive the Schr\"odinger-type equation.
In this case, our interesting region of $r\in[r_+,\infty]$ could be mapped into the whole region of  $r_*\in[-\infty,\infty]$,
which implies that the inner region of  $r<r_+$ is irrelevant to analyzing the superradiant stability.
Then,  the radial equation (\ref{wave-rad}) takes a form of
the Schr\"odinger-type equation when setting $\Psi_{lm}(r)=\sqrt{a^2+r^2} R_{lm}(r)$
\begin{equation}
\frac{d^2\Psi_{lm}(r_*)}{dr_*^2}+\Big[\omega^2-V_{KN}(r)\Big]\Psi_{lm}(r_*)=0, \label{sch-eq}
\end{equation}
where  the well-defined   potential $V_{KN}(r)$ is found to be~\cite{Benone:2014ssa}
\begin{eqnarray}
V_{KN}(r)=\omega^2&-&\frac{3\Delta^2r^2}{(a^2+r^2)^4}+\frac{\Delta[\Delta+2r(r-M)]}{(a^2+r^2)^3} \nonumber \\
&+&\frac{\Delta \mu^2}{a^2+r^2}-\Big[\omega-\frac{am}{a^2+r^2}-\frac{qQr}{a^2+r^2}\Big]^2\nonumber \\
&-&\frac{\Delta}{(a^2+r^2)^2}\Big[2am\omega -\lambda_{lm}\Big] \label{c-pot}.
\end{eqnarray}
Here we observe that  all $\Delta$ are located at the numerators, while all `$a^2+r^2$' appear in the denominators, in compared to the location of $\Delta(a^2+r^2)$ in denominators (numerators) for $V(r)$ in (\ref{f-pot}). This is because we use $R_{lm}(r)= \Psi_{lm}(r)/\sqrt{a^2+r^2}$ to derive $V_{KN}(r)$, whereas  $R_{lm}(r)= \Psi_{lm}(r)/\sqrt{\Delta}$ is used to  derive $V(r)$. In other words, we use the former to find out the Schr\"odinger-type equation (\ref{sch-eq}) written by $r_*$, while the latter is necessary to make $\Psi'_{lm}(r)$-term absent  in Eq.(\ref{wave-new}) written by $r$.  So, different choosing  $R_{lm}(r)$ makes different overall scale factor in the potential.

Replacing  $\lambda_{lm}$ by $\tilde{\lambda}_{lm}+a^2(\omega^2-\mu^2)$, one finds that $\omega^2-V_{KN}(r)$ =$V_{\omega lm}(r) $ in Ref.~\cite{Benone:2019all}.
It is noted that for $Q=0$, $V_{KN}(r)$ reduces to the potential $V_{K}(r)$  around the Kerr black hole~\cite{Zouros:1979iw} whose angular equation takes the form
\begin{eqnarray}
&& \frac{1}{\sin \theta}\partial_{\theta}\Big(
\sin \theta
\partial_{\theta} S_{\ell m}(\theta) \Big )+ \left [\tilde{\lambda}_{lm}+ a^2 (\omega^2-\mu^2) \cos^2
{\theta}-\frac{m^2}{\sin ^2{\theta}} \right ]S_{l m}(\theta) =0,
\label{wave-ang2}
\end{eqnarray}
which implies a relation of $\lambda_{lm}=\tilde{\lambda}_{lm}+a^2(\omega^2-\mu^2)(\lambda_{lm}<\tilde{\lambda}_{lm})$ with $\tilde{\lambda}_{lm}\sim l(l+1)+\cdots$. In this case, the last line of Eq.(\ref{c-pot}) is replaced  by
\begin{equation}
-\frac{\Delta}{(a^2+r^2)^2}\Big[2am\omega-a^2(\omega^2-\mu^2) -\tilde{\lambda}_{lm}\Big] \label{last-line}
\end{equation}
which  leads to potentials found in Refs.~\cite{Zouros:1979iw,Dolan:2007mj,Arvanitaki:2010sy,Konoplya:2011qq} for studying the superradiant instability.
In the non-rotating limit of $a\to0$, we could recover the scalar potential   from Eqs.(\ref{c-pot}) and (\ref{last-line})
when studying superradiance in the RN black hole spacetimes under a charged massive scalar propagation~\cite{Herdeiro:2013pia,Degollado:2013bha,DiMenza:2014vpa,Benone:2015bst}.

Before we proceed, let us explain a superradiant scattering by the KN black black holes.
We find two limits such  that $V_{KN}(r\to \infty)\to \mu^2$ and $V_{KN}(r\to r_+) \to \omega^2 -(\omega-\omega_c)^2$.
 The latter limit is obviously achieved because $\Delta \to 0$ as $r\to r_+$.
In this case, we have standard scattering forms of plane waves as~\cite{Brito:2015oca}
\begin{eqnarray}
\Psi_{lm}&\sim& {\cal I} e^{-i\sqrt{\omega^2-\mu^2} r_*}(\leftarrow)+{\cal R}e^{+i\sqrt{\omega^2-\mu^2} r_*}(\rightarrow),\quad r_*\to +\infty(r\to \infty) , \label{asymp1}\\
\Psi_{lm}&\sim& {\cal T} e^{-i(\omega-\omega_c) r_*}(\leftarrow),\quad r_*\to -\infty(r\to r_+) \label{asymp2}
\end{eqnarray}
with the ${\cal T}({\cal R})$ the transmission (reflection) amplitudes. The Wrongskian $W(\Psi,\Psi^*) $ of the complex conjugate solutions $\Psi$ and $\Psi^*$
satisfies
\begin{equation}
i \frac{d}{dr_*} W(\Psi,\Psi^*)=0,
\end{equation}
which implies that
\begin{equation}
|{\cal R}|^2=|{\cal I}|^2-\frac{\omega-\omega_c}{\sqrt{\omega^2-\mu^2}}|{\cal T}|^2.
\end{equation}
This means that only waves with $\omega>\mu$ propagate to infinity and the superradiant scattering occurs ($\rightarrow,~|{\cal R}|^2>|{\cal I}|^2$) whenever $\omega<\omega_c$.
Actually, the superradiance  is associated to  having  a negative absorption cross section~\cite{Benone:2019all}. However, it turned out that the scalar  absorption cross section is always positive for plane waves. For the KN black hole, the total absorption cross section becomes negative for co-rotating spherical waves at low frequencies.
The superradiance can occur for massive scalar waves as long as the superradiance  condition
\begin{equation}
\omega<\omega_c   \label{s-cond}
\end{equation}
is satisfied.
Now, we wish  to describe the superradiant instability briefly.
The interaction between  a rotating  black hole and a massive scalar field will prevent low frequency modes with $\omega<\mu$ from escaping to spatial infinity.
It is well known that if  a massive scalar  with mass $\mu$ is scattered off by a
rotating black hole, then for $\omega<\mu$, the superradiance with $\omega<\omega_c$  might have  unstable modes because the mass term  works effectively as a reflecting mirror. In this case, a potential shape between ergo-region and mirror-region has a local maximum (potential barrier) as well as a local minimum (trapping well) away  from the outer horizon which generates a secondary reflection of the wave that was reflected from the potential barrier~\cite{Zouros:1979iw,Arvanitaki:2010sy,Konoplya:2011qq}. The secondary reflected wave will be reflected again in the far region. Since each scattering off the barrier in the superradiant region increases the amplitude of the wave, the process of reflections will continue with the increased energy of waves, leading to an instability. This implies  a quasi-bound state which is an approximate energy eigenstate localized in the scattering region.
When an incident wave enters the scattering region with the right energy $E=\omega^2$, it spends a long time trapped in the quasi-bound state before eventually escaping back to infinity.
The corresponding boundary conditions imply an exponentially decaying wave away from trapping well and a purely outgoing wave near the outer horizon:
\begin{eqnarray}
\Psi&\sim& e^{-\sqrt{\mu^2-\omega^2}r}({\rm bound-state}),\quad r_* \to \infty(r\to \infty), \label{rad-sol1} \\
\Psi&\sim& e^{-i(\omega-\omega_c)r_*}(\rightarrow:~{\rm superradiance}),\quad r_* \to -\infty(r\to r_+).  \label{rad-sol2}
\end{eqnarray}
Considering a time-dependance $e^{-i \omega t}$ in (\ref{sep}), the outgoing wave could be achieved when satisfying  $\omega<\omega_c$.
To obtain a decaying mode at spatial infinity, one needs to have a bound-state condition of $\omega<\mu$. Therefore, the conditions for the superradiant instability
are given by
\begin{equation}
\omega<\omega_c, \quad \omega<\mu. \label{si-cond}
\end{equation}
Importantly, the other (necessary) condition comes from the existence of  a local minimum (a positive trapping well).
The stable/unstable nature is selected  by a shape of the potential.
If there is no trapping well, it corresponds to a superradiant stability.
As is shown in (Right) Fig. 2, $V_{KN}(r\to r_+)\simeq\omega^2-(\omega-\omega_c)$ is $-0.013$, while $V(r\to r_+)=-1.324\times 10^{13}$ is singular [see  (Left) Fig. 1]. The latter is inappropriate  for imposing the superradiance condition (\ref{rad-sol2}) near  the horizon. Here, we use the same parameters for obtaining  $V_{KN}(r)$ and $V(r)$. Also, we find a significant difference between   $V_{KN}(r\to r_+)=0.06$ and $V(r\to r_+)=-4.98\times 10^{31}$ in the Kerr black hole when setting $q=Q=0$ with the same parameters. Thus, it is conjectured that  there is no parameter space for which $V(r)$ matches $V_{KN}(r)$ closely because their dependance of $\Delta$ is quite different.

\begin{figure*}[t!]
   \centering
  \includegraphics{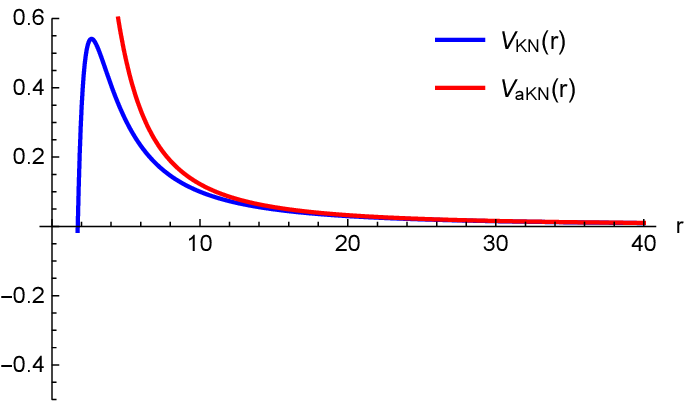}
  \hfill%
  \includegraphics{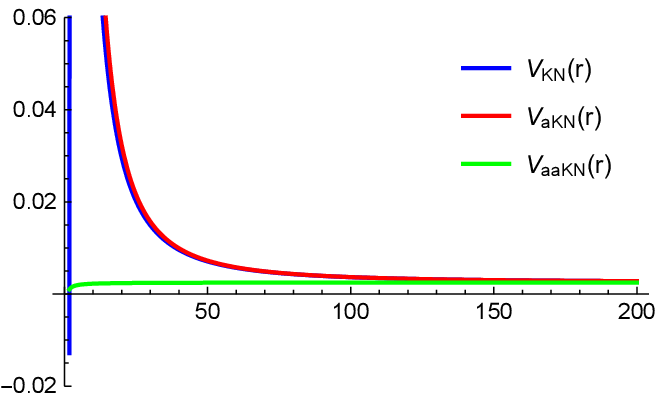}
\caption{(Left) Potential  without trapping well $V_{KN}(r)$ as function of $r\in[r_+=1.741,40]$ with $M=1,Q=0.6,\omega=0.02,a=0.3,m=1,q=0.1,\lambda_{lm}=12,\mu=0.05$.
  $V_{aKN}(r)$ mimics $V_{KN}(r)$ for $r\ge 20$.
(Right) $V_{KN}(r_+)$ is finite ($-0.013$) like as $\omega^2-(\omega-\omega_c)^2=-0.012$. $V_{aaKN}(r)$ could  imitate $V_{KN}(r)$  for $r\ge150$. We check the condition of  $\omega<\omega_c=0.13$ and $\omega<\mu$  in  Eq. (\ref{si-cond}). }
\end{figure*}
It seems that Fig. 2 corresponds to a  superradiantly stable potential  because we could not find a trapping well  for $\omega<\omega_c=0.129$ and $\omega<\mu$.
In this case, we observe  that $\lim_{r\to 9225.94}V_{KN}(r)\to [\searrow\mu^2]$.
Here, we find that a tiny  well is located at $r=9225.94$ but it does not affect the superadiant stability.

To find out the existence of  a local minimum, we  consider   $V_{aKN}(r)$ (when expanding $V_{KN}(r)$ for  large $r$) given by
\begin{equation}
V_{ KN}(r\to\infty) \to V_{aKN}(r) +{\cal O}\Big(\frac{1}{r^3}\Big),\label{asymKN}
\end{equation}
where
\begin{equation}
V_{aKN}(r)=\mu^2-\frac{2(M \mu^2-q Q \omega)}{r}+\frac{\lambda_{lm}+Q^2(\mu^2-q^2)}{r^2}. \label{apKN}
\end{equation}
Here, the third term plays an important role in making a trapping well  because $\lambda_{lm}$ may take a large value.
In this case, we point out that the condition for (no) trapping well is given by $(V'_{aKN}(r)<0)~V'_{aKN}(r)>0$.
From (Right) Fig. 2, we know that $V_{aKN}(r)$ mimics  $V_{KN}(r)$ well for large  $r\ge 20$ and $V'_{aKN}(r)<0$ implies no trapping well.

On the other and, $V_{aaKN}(r)$  defined  through (when expanding $V_{KN}(r)$ for very large $r$)
\begin{equation}
V_{ KN}(r\to \infty) \to V_{aaKN}(r) +{\cal O}\Big(\frac{1}{r^2}\Big),\label{asymvKN}
\end{equation}
where
\begin{equation}
V_{aaKN}(r)=\mu^2-\frac{2(M \mu^2-q Q \omega)}{r} \label{aapKN}
\end{equation}
could  represent $V_{KN}(r)$ for very large   $r\ge 150$.  This implies that $V_{aaKN}(r)$ is not sufficient to derive the condition for no trapping well.
In this case, we stress that it is  dangerous  to derive a condition for no trapping well from $V_{aaKN}(r)$ only.
If the condition of no trapping well  is given by  $V_{aaKN}'(r)<0$, it implies a bound on $\omega$ as
\begin{equation}
\omega>\omega_d~{\rm with}~\quad \omega_d=\frac{M\mu^2}{qQ}. \label{no-twc}
\end{equation}
\begin{figure*}[t!]
   \centering
  \includegraphics{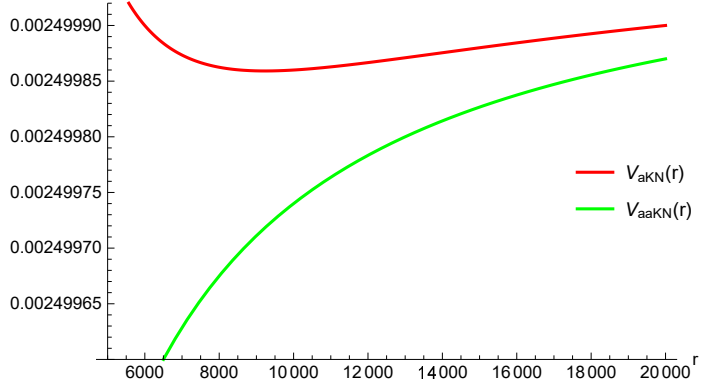}
\caption{Asymptotic forms of  $V_{aKN}(r)$ and $V_{aaKN}(r)$. $V_{aKN}(r)$ shows a tiny well located at $r=9228.69$ and it approaches $V_{aaKN}(r)$ for $r>9228.69$.
All coefficients go together with Fig. 3.  }
\end{figure*}
However, we point out that  this bound is not satisfied even for a superradiantly stable potential in  Fig. 2 because of $\omega(=0.02)<\omega_d(=0.042)$.
As is shown in Fig. 3, $V_{aKN}(r)$ has a tiny  well at $r=9228.69$ and it approaches $V_{aaKN}(r)$ for $r>9228.69$.
This explains why $\omega>\omega_d$ in Eq. (\ref{no-twc}) does not hold at asymptotic region.

At this stage, it is worth noting that $V_{aKN}(r)$ with $a=0$ and $\lambda_{lm}=l(l+1)$ and $V_{aaKN}(r)$ could be found exactly from the scalar potential obtained  when charged massive scalar modes are impinging on the RN black holes~\cite{Herdeiro:2013pia,Degollado:2013bha,DiMenza:2014vpa,Benone:2015bst}.
\begin{figure*}[t!]
   \centering
  \includegraphics{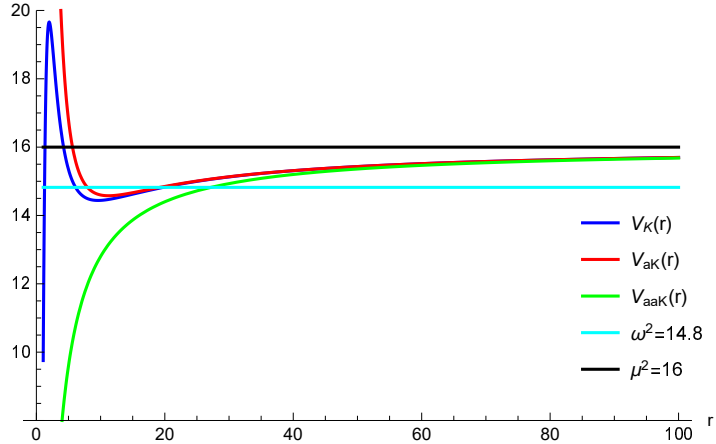}
\caption{Positive Potential $V_{K}(r)=V^{Q=0}_{KN}(r)$ and its asymptotic potential $V_{aK}(r)=V^{Q=0}_{aKN}(r)$ around Kerr background as function of $r\in[r_+=1.063,100]$ with $M=1,Q=0,\omega=3.85,a=0.998,m=13,q=0,\lambda_{lm}=180,\mu=4$.
The potential is finite (9.754) at $r=r_+$  and it has a trapping well (local minimum) at $r=9.6$. $V_{aK}(r)$ mimics  $V_{K}(r)$ well for $r\ge 20$, while  $V_{aaK}(r)$ mimics  $V_{K}(r)$ for $r\ge 50$.}
\end{figure*}

Now, we are in a position to introduce a  potential with trapping well which is a necessary condition for  the superradiant instability.
Firstly, we consider a  massive scalar propagation  on the Kerr black hole background.
In this case, we find a positive potential shown in Fig. 4, which involves a  trapping well (local minimum) located at $r=9.6$ and  indicates $V'_{aK}(r)>0$ for large $r$. Plotting $V_K$ in terms of the tortoise coordinate $r_*$,
$V_K$ is invariant in the depth but its near-horizon region  is stretched from $-\infty$ (the horizon) to 0.
In this case,  one may recover
Fig. 2 in Ref.~\cite{Zouros:1979iw} with the regions I (ergo-region: near-horizon), II (barrier-region), III (well-region), and IV (mirror-region: far-region) [Fig. 7 in Ref.~\cite{Arvanitaki:2010sy}
and Fig. 15 in Ref.~\cite{Konoplya:2011qq}]. Three (II, III, IV) are essential for realizing superradiant instability and  these regions are divided by imposing the turning point  condition of $V_{K}=\omega^2$.
Here, we note that  there are quasi-bound states but  there are no  genuine bound states because the potential is purely repulsive ($V_{K}(r)>0$, everywhere).
\begin{figure*}[t!]
   \centering
  \includegraphics{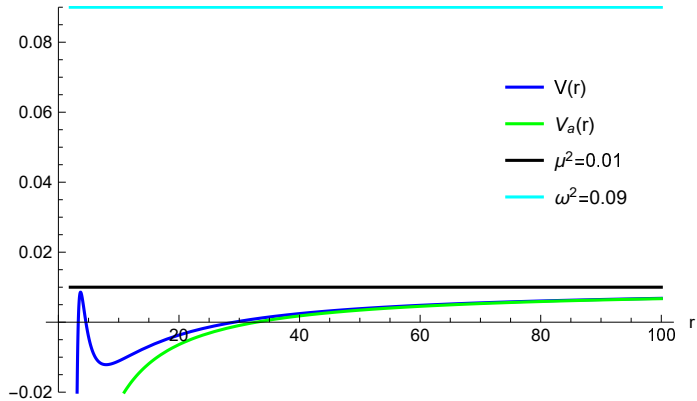}
\caption{ Potential $V(r)$ as function of $r$ with $M=1,Q=0.2,\omega=0.3,a=0.1,m=1,q=0.1,\lambda_{lm}=2,\mu=0.1$  for $r\in[r_+=1.975,100]$.
The potential is negatively singular  at $r=r_+$ with a negative well.  $V_a(r)$ mimics $V(r)$   for $r\ge 40$ but $V(r)<\omega^2$ and $\omega>\mu$.}
\end{figure*}

In  case of $V(r)$ in Eq.(\ref{f-pot}), we display a case of potential with trapping well  mentioned in Ref.~\cite{Xu:2020fgq} in Fig. 5.
However, it is not suitable for  representing  an example for  the superradiant instability because the bound-state condition ($\omega<\mu$) is not satisfied and there is no turning point (because of $V\ll\omega^2$). Furthermore, $V(r)$ is negatively singular ($-5.16\times 10^6$) at $r=r_+$ and it has a negative trapping well.

Considering  a charged massive scalar on the KN black hole with $a=0.998$ (rapidly rotating black hole),
 we find a positive potential (see Fig. 6), which involves a trapping well (local minimum located at $r=9.61$) and shows $V'_{aKN}(r)>0$ for large $r$.  We note that $\lim_{r\to \infty}V_{KN}(r)\to [\nearrow\mu^2]$ for trapping well, compared to  $\lim_{r\to 9225.94}V_{KN}(r)\to [\searrow\mu^2]$ for no trapping well in Fig. 2.  This induces a quasi-bound state,  leading to  the superradiant instability.
Comparing  Fig. 6 with Fig. 2, it is worth noting that $V_{aKN}(r)$ includes  a trapping well, but  $V_{aaKN}(r)$ does not include  a trapping well.
\begin{figure*}[t!]
   \centering
  \includegraphics{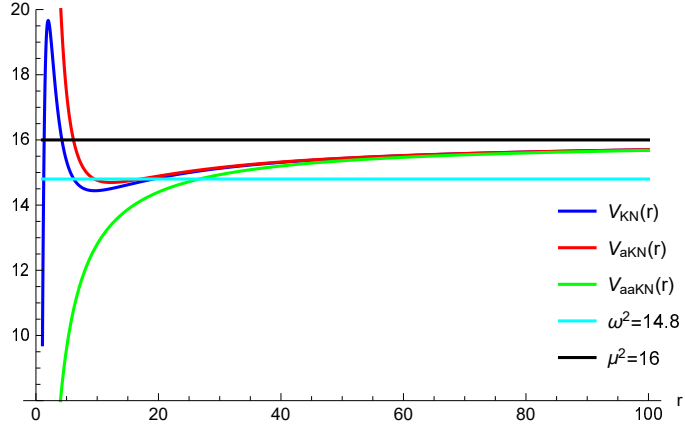}
\caption{Positive potential $V_{KN}(r)$  and its asymptotic potential $V_{aKN}(r)$ as functions of $r\in[r_+=1.062,100]$ with $M=1,Q=0.01,\omega=3.85,a=0.998,m=13,q=0.2,\lambda_{lm}=180,\mu=4$.
$V_{KN}(r)$ is finite (9.728) at $r=r_+$ and it has a trapping well located at $r=9.61$. Also, $V_{aKN}(r)$ imitates $V_{KN}(r)$ well for $r\ge 20$ whereas  $V_{aaKN}(r)$ mimics  $V_{KN}(r)$ for $r\ge 50$.
We note   $\omega<\omega_c=5.6398$ and $\omega<\mu$ as two conditions for superradiant instability. }
\end{figure*}
This implies that it is not valid to derive a condition for no trapping well directly from $V_{aaKN}(r)$. In this case, we note that the condition for no trapping well
[Eq.(\ref{no-twc})] violates because of $\omega(=3.85)<\omega_d=8000$.

At this stage, we wish to mention the relation between tortoise coordinate and trapping well.
We know  from Eq. (\ref{c-pot}) that the first line without $\omega^2$ represents the effect of introducing the tortoise coordinate $r_*$, while the last two lines come from $-U(r)/(a^2+r^2)^2$ in Eq. (\ref{u-pot}). Actually, the latter determines  the asymptotic potential $V_{aKN}(r)$ in Eq. (\ref{apKN}) completely whose
third term plays an important role in making a trapping well for a large $\lambda_{lm}$. This means that $V'_{aKN}(r)$ in assessing the trapping well is not affected by introducing the tortoise coordinate.  Since $V'_{aKN}(r)>0$ for large $r$ implies the presence of a trapping well, introducing the tortoise coordinate $r_*$ does not affect the presence of a trapping well. However, let us compare $V(r)$  in Fig. 5 with $V_{KN}(r)$ in Fig. 6.  Even though they have a  well, the former does not satisfy the bound-state condition ($\omega< \mu$), it is negatively singular at the horizon, and it has a negative trapping  well.

To obtain  stationary bound-state resonances, one has two conditions of $\omega=\omega_c$ and $\omega<\mu$.
They correspond to  marginally stable modes of the scalar field with Im[$\omega$]=0, leading to  scalar clouds.
In fact, such stationary resonances saturate the superradiance condition Eq.(\ref{s-cond}). However, even for weakly bound stationary resonances ($\omega=\omega_c\le\mu$), there exist several distinct physical regimes~\cite{Hod:2014baa}.
We display a potential of stationary bound-state resonances in Fig. 7, which does not include a trapping well~\cite{Benone:2014ssa}. This  potential is similar to Fig. 2 except $\omega=\omega_c$.
\begin{figure*}[t!]
   \centering
  \includegraphics{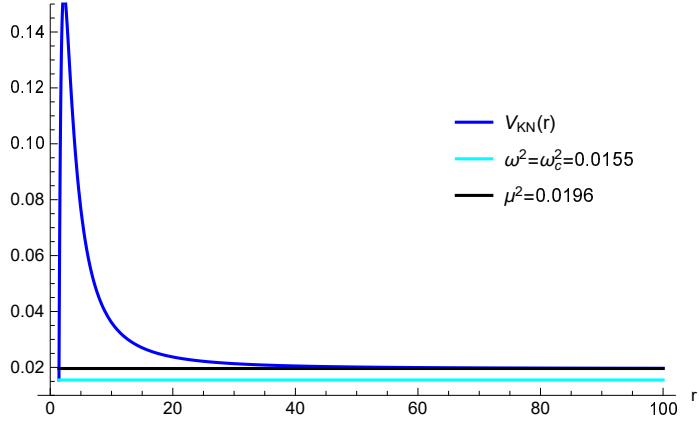}
\caption{Stationary resonance  potential $V_{KN}(r)$   as function of $r\in[r_+ =1.4243,100]$ with $M=1,Q=0.9,\omega=0.1245,a=0.1,m=1,q=0.12,\lambda_{lm}=2,\mu=0.14$.
$V_{KN}(r\to r_+)$ is finite ($\omega^2=0.0155$).
We note   ``$\omega=\omega_c$ and $\omega<\mu$" to meet the condition for   stationary bound-state resonances. }
\end{figure*}

Finally, we summarize   four cases  for a massive scalar propagating around the KN black holes:\\
(i) superradiant scattering  $\to$ $\omega<\omega_c$ and $\omega>\mu$. \\
(ii) stationary bound-state resonances $\to$ $\omega=\omega_c$ and $\omega<\mu$.\\
(iii) superradiant instability $\to$ $\omega<\omega_c$ and  $\omega<\mu$  with  a positive trapping well.\\
(iv) superradiant stability $\to$ $\omega<\omega_c$ and $\omega<\mu$  without a positive trapping well. \\

\section{Scalar waves in the  far-region}

It is important to know the scalar wave forms in the  far-region  to distinguish between trapping well and no trapping well.
In this direction, we wish to derive wave functions  in the far-region.

In the far-region where we have $r_*\sim r$ [$\Psi_{lm}(r)\sim rR_{lm}(r)$], we obtain an  equation from Eqs. (\ref{sch-eq}) and (\ref{apKN}) as
\begin{equation}
\Big[\frac{d^2}{dr^2}+\omega^2-V_{aKN}(r)\Big]\Psi_{lm}=0
\end{equation}
whose solution is given by the  confluent  Hypergeometric function $U[a,b,x]$ as
\begin{eqnarray}
\Psi_{lm}&=&c_1 e^{-\sqrt{\mu^2-\omega^2}r} \Big(2\sqrt{\mu^2-\omega^2} r\Big)^{\frac{1}{2}+\tilde{m}} \nonumber \\
 &\times & U\Big[\frac{1+2\tilde{m}}{2}-\frac{M\mu^2-qQ\omega}{\sqrt{\mu^2-\omega^2}},1+2\tilde{m},2\sqrt{\mu^2-\omega^2} r\Big]. \label{wavef-1}
\end{eqnarray}
with
\begin{equation}
\tilde{m}=\frac{1}{2}\sqrt{1+4[\lambda_{lm}+Q^2(\mu^2-q^2)]}.
\end{equation}
 Here we find an asymptotic  bound-state of  $e^{-\sqrt{\mu^2-\omega^2}r}$ appeared  in (\ref{rad-sol1}).

If one uses $V_{aaKN}(r)$ in Eq.(\ref{aapKN}) with $\tilde{m}=1/2$, its solution is given by
\begin{eqnarray}
\Psi_{lm}=c_2 e^{-\sqrt{\mu^2-\omega^2}r}\Big(2\sqrt{\mu^2-\omega^2} r\Big)~ U\Big[1-\frac{M\mu^2-qQ\omega}{\sqrt{\mu^2-\omega^2}},2,2\sqrt{\mu^2-\omega^2} r\Big].\label{wavef-2}
\end{eqnarray}
\begin{figure*}[t!]
   \centering
  \includegraphics{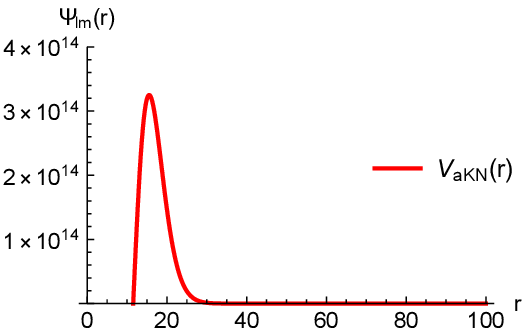}
  \hfill%
  \includegraphics{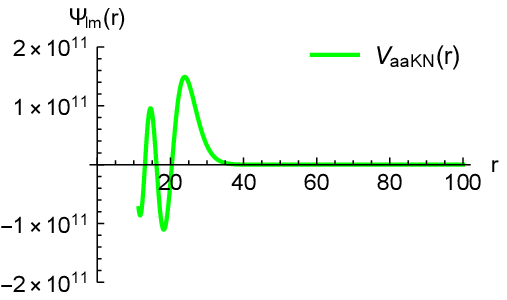}
\caption{Radial modes $\Psi_{lm}$ as function of $r\in[11.3,100]$ with trapping well: (Left) quasi-bound state with $V_{aKN}(r)$ and (Right) oscillation with $V_{aaKN}(r)$.
 All parameters  go together with Fig. 6.  }
\end{figure*}
Let us observe radial modes  $\Psi_{lm}(r)$ for the case of superradiant instability (see Fig. 6).
As is shown in (Left) Fig. 8, Eq.(\ref{wavef-1}) shows a  quasi-bound state followed by exponentially decaying mode, whereas  Eq.(\ref{wavef-2})   indicates an oscillation followed by exponentially decaying mode [see (Right) Fig. 8].
\begin{figure*}[t!]
   \centering
  \includegraphics{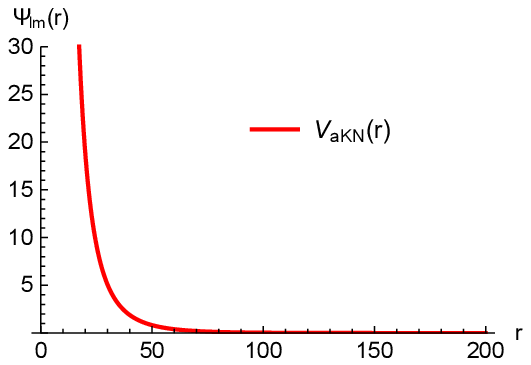}
  \hfill%
  \includegraphics{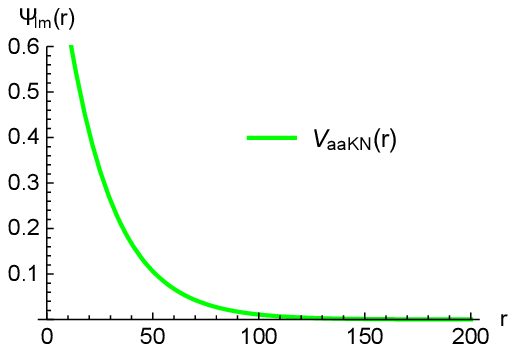}
\caption{ Radial modes $\Psi_{lm}$ as function of $r\in [12,100]$ without trapping well: (Left) exponentially decaying function with $V_{aKN}(r)$ and (Right) exponentially decaying function with $V_{aaKN}(r)$.  All parameters  go together with Fig. 2.  }
\end{figure*}
Contrastively, we consider radial modes for the potential  without trapping well (shown in Fig. 2), implying superradiant stability.
As is shown in Fig. 9, Eqs.(\ref{wavef-1}) and (\ref{wavef-2})   show  exponentially decaying modes, describing bound states.

 From (Left) Fig. 8 and (Left) Fig. 9, we observe  a difference between quasi-bound state and bound state.
 It depends on the sign of the first argument $a$ in the confluent hypergeometric function $U[a,b,x]$ in Eq.(\ref{wavef-1}).
 As is shown in (Left) Fig. 10, the superradiant instability (trapping well) could be achieved whenever $a$ is negative as
 \begin{equation}
 a<0 \to \quad \frac{M\mu^2-qQ\omega}{\sqrt{\mu^2-\omega^2}}>\tilde{m}+\frac{1}{2} \label{sin-cond}
 \end{equation}
 together with $\omega<\omega_c$ and $\omega<\mu$.
 On the other hand, from (Right) Fig. 10 we have  the superradiant stability (no trapping well) for  positive $a$ as
 \begin{equation}
 a>0 \to \quad \frac{M\mu^2-qQ\omega}{\sqrt{\mu^2-\omega^2}}<\tilde{m}+\frac{1}{2}  \label{sst-cond}
 \end{equation}
 together with $\omega<\omega_c$ and $\omega<\mu$.
 \begin{figure*}[t!]
   \centering
  \includegraphics{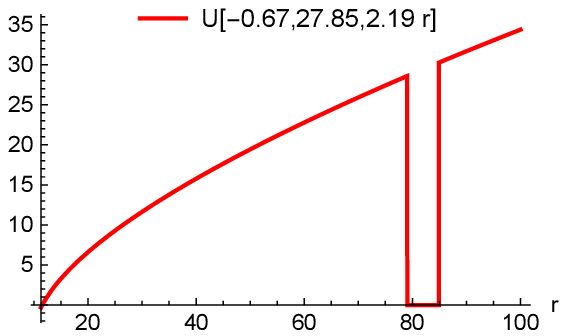}
  \hfill%
  \includegraphics{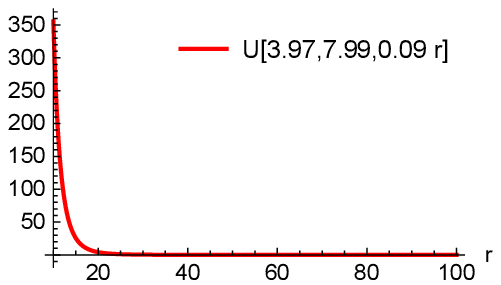}
\caption{ (Left) Confluent hypergeometric function $U[-0.8,27.8;2.17 r]$ for quasibound state is an increasing function of $r$.
(Right) Confluent hypergeometric function $U[3.97,7.99;0.09 r]$ for bound state is a decreasing function of $r$.  }
\end{figure*}

\begin{figure*}[t!]
   \centering
  \includegraphics{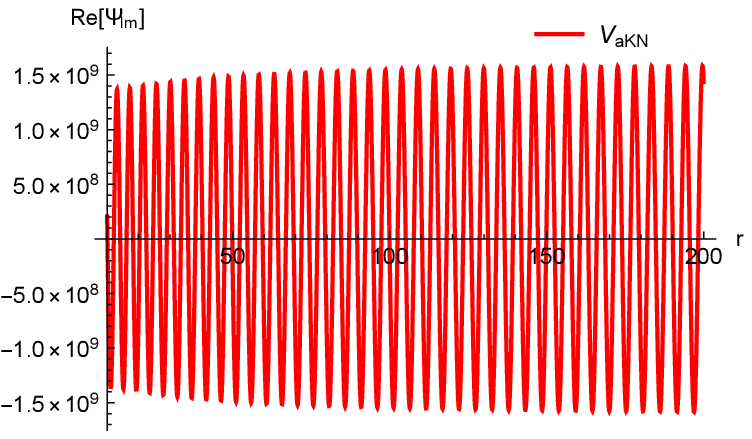}
  \hfill%
  \includegraphics{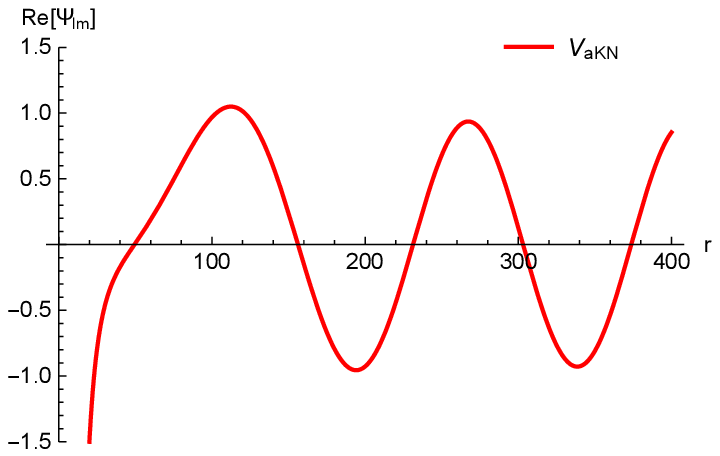}
\caption{(Left) Real part of  $\Psi_{lm}(r)$ as function of $r\in [10,200]$  where all parameters go together with Fig. 7 except $\omega=4>\mu=3.85$. (Right)  Real part of  $\Psi_{lm}(r)$ as function of $r\in [10,400]$ where all parameters go together with Fig. 3 except $\omega=0.05>\mu=0.02$.  }
\end{figure*}
On the other hand, it is interesting to  observe  superradiant scattering pictures for $\omega>\mu$.  As is shown in Fig. 11, they are plane waves but their differences appear in the wave number and amplitude when adopting $V_{aKN}(r)$. These pictures are compared to (Left) Fig. 8 and (Left) Fig. 9.
However, it is known that the scalar  absorption cross section is always positive for plane waves. For the KN black hole, the total absorption cross section becomes negative for co-rotating spherical waves at low frequencies~\cite{Benone:2019all}.

\section{Discussions}

It was reported  that the superradiant stability of the KN black hole can be achieved if $qQ/\mu M> 1$ and $r_-/r_+\le 1/3$  are satisfied  when analyzing the potential $V(r)$ in Eq.(\ref{f-pot})~\cite{Xu:2020fgq}. Honestly, one could not regard $V(r)$ as a correct potential because it is negatively singular at the outer horizon as well as  it was derived from  a radial wave equation (\ref{wave-rad}) without introducing the tortoise coordinate $r_*$. We note that  the potential $V(r)$ was obtained just by imposing on the disappearance of $\Psi'_{lm}$ in Eq.(\ref{wave-new}) from  $\Psi_{lm}(r)=\sqrt{\Delta} R_{lm}(r)$ with $r$. It is worth noting  that an opposite bound of $qQ/\mu M< 1$ was firstly denoted as  a condition for getting a trapping well (superradiant instability) around the KN black hole~\cite{Furuhashi:2004jk}.  However, this condition is not satisfied simultaneously  when imposing the superradiance condition ($\omega<\omega_c$) and thus, it is considered as a condition for bound states~\cite{Degollado:2013bha}. Also, we note that  their effective potential $V_{\rm eff}(r)$ belongs to  a shortened form because  $\Psi_{\ell m}(r)=rR_{\ell m}(r)$ and a modified tortoise coordinate $x$ with $dx=r^2dr/\Delta$ are used to derive it.

In this work, we have found a correct potential $V_{KN}(r)$ in  Eq.(\ref{c-pot}) from $\Psi_{lm}(r)=\sqrt{a^2+r^2} R_{lm}(r)$ with $r_*$ to discuss the condition for  superradiant stability (no trapping well).
 To show the existence of  a trapping well  is  a necessary condition for the  superradiant instability because superradiant modes are localized in the trapping well~\cite{Arvanitaki:2010sy}. If there is no trapping well, it means the superradiant stability.
For superradiant stability, one needs to check the condition for   no trapping well, in addition to two boundary conditions: $\omega<\omega_c$ and ($\omega<\mu$).  Actually, the condition for no trapping well is given by $V'_{aKN}(r)<0$.  However, it is not easy to derive any analytic condition for no trapping well from $V'_{aKN}(r)<0$ directly.
As is shown in Fig. 8 (superradiant instability), Eq.(\ref{wavef-1}) with $V_{aKN}(r)$  shows a  quasi-bound state followed by exponentially decaying mode, whereas  Eq.(\ref{wavef-2}) with $V_{aaKN}(r)$    indicates an oscillation followed by exponentially decaying mode.
On the other hand, for the case of superradiant stability (Fig. 9), Eqs.(\ref{wavef-1}) and (\ref{wavef-2})   show  exponentially decaying modes only.
From the observation of its asymptotic scalar function $U[a,b,x]$ in Fig. 10 based on $V_{aKN}(r)$, we find  Eq.(\ref{sin-cond}) for a condition of  superradiant instability and  Eq.(\ref{sst-cond}) as a condition of  superradiant stability.

\begin{figure*}[t!]
   \centering
  \includegraphics{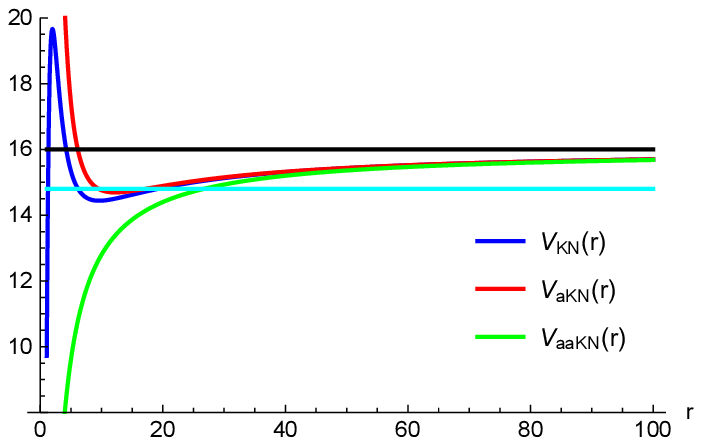}
  \hfill%
  \includegraphics{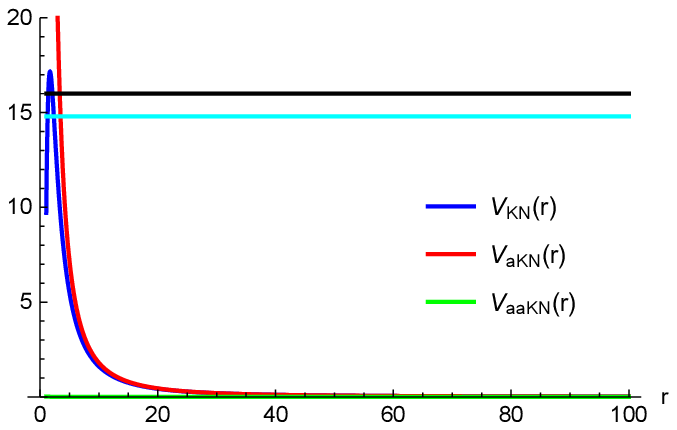}
\caption{(Left) Potential $V_{KN}(r)$ and its asymptotic potentials $V_{aKN}(r)$ and $V_{aaKN}(r)$ as functions of $r\in[r_+=1.062,100]$ with $M=1,Q=0.01,\omega=3.85,a=0.998,m=13,q=0,\lambda_{lm}=180,\mu=4$.
$V_{KN}(r\to r_+)$  is 9.72 and a trapping well is located at $r=9.6$. We note   $\omega<\omega_c=6.11$ and $\omega<\mu$ as the condition for superradiant instability.
(Right) Potential $V_{KN}(r)$ and its asymptotic potentials $V_{aKN}(r)$ and $V_{aaKN}(r)$ as functions of $r\in[r_+=1.062,100]$ with $M=1,Q=0.01,\omega=3.85,a=0.998,m=13,q=0.2,\lambda_{lm}=180,\mu=0$.  There is no trapping well because of $\mu=0$ and $\omega<\mu$ is not satisfied.}
\end{figure*}

Finally, we wish to discuss  the limitation on  superradiance and superradiant instability.  As was shown in~\cite{Brito:2015oca}, black hole superradiance is a radiation enhancement process that allows for energy extraction from the black holes at the classical level. This process is available from  the static charged black hole, the rotating black hole, the charged rotating black hole, and the analogue black hole geometries. On the other hand, Press and Teukolsky~\cite{Press:1972zz} have  suggested the `rotating black hole-mirror bomb' idea: if the superradiance emerging from a perturbed black hole
were reflected back onto the rotating black hole by a spherical mirror, an initially small perturbation could be made to grow without bound~\cite{Cardoso:2004nk}.
This superradiant instability is caused by either the mirror (artificial wall) or the cavity (AdS background).
The reflection will occur naturally if a perturbed bosonic field has a rest mass $\mu$~\cite{Damour:1976kh}.
The superradiant instability is surely possible to occur in the KN black hole (see Fig. 6 for its charged massive scalar potential with trapping well) and in the Kerr  black hole (see Fig. 4 for its massive scalar potential with trapping well).  We could have the superradiant instability for a massive scalar with $q=0$ around  the KN black hole [see (Left) Fig. 12 with trapping well], whereas it is hard to have the superradiant instability for  a charged scalar with $\mu=0$ because there is no mirror [see (Right) Fig. 12 without trapping well]. In addition, it is worth noting  that the superradiant instability is not  found from a charged  massive scalar around the RN black holes.
However, the superradiant instability of a charged massive scalar  could be obtained if the RN black hole is enclosed in a cavity~\cite{Herdeiro:2013pia,Degollado:2013bha}. This is called the charged black hole-mirror bomb, which is a spherically symmetric analogue of the rotating black hole-mirror bomb.

 \vspace{0.5cm}

{\bf Acknowledgments}
 \vspace{0.5cm}

This work was supported by a grant from Inje University for the Research in 2021 (20210040).

\end{document}